# A Class of Orthogonal Sequences

Subhash Kak
November 26, 2013

**Abstract**
This paper presents a class of random orthogonal sequences associated with the number theoretic Hilbert transform. We present a constructive procedure for finding the random sequences for different modulus values. These random sequences have autocorrelation function that is zero everywhere excepting at the origin. These sequences may be used as keys and in other cryptography applications.

**Introduction**
Near-orthogonal random sequences, such as PN sequences, are used as keys in spread spectrum for communications security [1],[2]. These sequences can also be used for providing security in wireless and sensor network applications. PN sequences are binary and in advanced systems they can be replaced by non-binary sequences for which perfectly orthogonal sequences based on circulant matrices which can be designed. Each of these orthogonal sequences can serve as a portal for information. One can also design groups of sequences that are correlated amongst themselves and orthogonal to other sequence groups; these can be used to aggregate users who can communicate freely amongst themselves.

Circulant matrices are already used in the mix columns of the Advanced Encryption Standard. Normally they are not viewed from the perspective of orthogonality, a perspective that becomes obvious once we see their relationship to the number theoretic Hilbert transform (NHT) [3]. The circulant matrix at the basis of this transform is derived from the standard discrete Hilbert transform [4]. The NHT circulant matrix (which is self-orthogonal with respect to a prime modulus) has alternating entries in each row of zero and non-zero numbers. When the NHT is viewed as a matrix with the 0 entries purged, its use as a random sequence generator becomes apparent. The discrete Hilbert transform has applications is a variety of areas of signal processing such as spectral analysis in 2-D reconstruction [4]-[13], multilayered computations [14],[15], and cryptography [16]-[18].

Here we present a constructive procedure to write down the NHT matrix of any size for this class and derive random orthogonal sequences from them. We will describe some properties that are satisfied by such orthogonal sequence sets. We will also show that they can be put in different sub-classes that can connect groups of users.

**NHT Circulant Matrices and Orthogonal Sequences**

Let the data block be the vector F, and the NHT transformed data block be the vector G. The NHT transform is the matrix N and the computations are with respect to the modulus p, a suitable prime. The inverse of the NHT matrix is $N^{-1} = N^T$ mod p. Therefore

$$G = NF \mod p, \text{ and}$$
$$F = N^T G \mod p \qquad (1)$$

It follows from the circulant property underling the mathematics that the product $NN^T$ performs various autocorrelation computations associated with the sequence in the first row of the matrix. This means that finding different NHT matrices also gives solutions for corresponding length discrete sequences that have ideal autocorrelation properties modulo the chosen p.

*Generator of NHT matrix:* The elements in the first row of the NHT matrix, that is the integers a, b, c, d, e, f, g, and so on, that alternate with 0s as the generator of the NHT matrix.

$$N = \begin{bmatrix} 0 & a & 0 & b & 0 & c & 0 & . & . & k \\ k & 0 & a & 0 & b & 0 & c & 0 & . & . \\ . & k & 0 & a & 0 & b & 0 & c & 0 & . \\ . & . & k & 0 & a & 0 & b & 0 & c & 0 \\ 0 & . & . & k & 0 & a & 0 & b & 0 & c \\ c & 0 & . & . & k & 0 & a & 0 & b & 0 \\ 0 & c & 0 & . & . & k & 0 & a & 0 & b \\ b & 0 & c & 0 & . & . & k & 0 & a & 0 \\ 0 & b & 0 & c & 0 & . & . & k & 0 & a \\ a & 0 & b & 0 & c & 0 & . & . & k & 0 \end{bmatrix} \mod p \qquad (2)$$

and *p* is an suitable value of the modulus, which is a prime number.

Now for ease of presentation, we purge the zeros in (2) and call the reduced matrix R, which will also satisfy the orthogonality property like N.

$$R = \begin{bmatrix} a_1 & a_2 & a_3 & . & a_M \\ a_M & a_1 & a_2 & a_3 & . \\ . & a_M & a_1 & a_2 & a_3 \\ a_3 & . & a_M & a_1 & a_2 \\ a_2 & a_3 & . & a_M & a_1 \end{bmatrix} \mod p \qquad (3)$$

The constraints on p emerge from the condition $RR^T=I$. The non-diagonal terms of the product (3) should be zero:

$$\sum_i a_i a_{i+j} = 0 \bmod p \quad \text{for } j \neq 0 \tag{4}$$

where $a_{M+i} = a_i$

The way to guarantee this is to choose the values of the entries in the circulant matrix (3) in such a way so that all the non-diagonal terms are the same (but different from the diagonal term) and then pick the value as the modulus.

Thus for M=5, the following two expressions must both be zero or equal so that we can pick this expression as the modulus:

$$\begin{aligned} a_1 a_5 + a_2 a_1 + a_3 a_2 + a_4 a_3 + a_5 a_4 \\ a_1 a_4 + a_2 a_5 + a_3 a_1 + a_4 a_2 + a_5 a_3 \end{aligned} \tag{5}$$

*Example 1.* Consider M=5 where all $a_i$ but $a_2$ are 1 and $a_2=b$. This leads to the condition $RR^T=kI$, where k is some constant. The reason why we don't insist on k being 1 is that as a random sequence modulo a prime, the degree of self-correlation is not an issue as it will vary based on what constant the equation has been multiplied with. This means that:

$$\begin{bmatrix} 1 & b & 1 & 1 & 1 \\ 1 & 1 & b & 1 & 1 \\ 1 & 1 & 1 & b & 1 \\ 1 & 1 & 1 & 1 & b \\ b & 1 & 1 & 1 & 1 \end{bmatrix} \begin{bmatrix} 1 & 1 & 1 & 1 & b \\ b & 1 & 1 & 1 & 1 \\ 1 & b & 1 & 1 & 1 \\ 1 & 1 & b & 1 & 1 \\ 1 & 1 & 1 & b & 1 \end{bmatrix} = kI \bmod p \tag{6}$$

We get another circulant matrix whose first row entries are:

$$b^2 + 4, 2b+3, 2b+3, 2b+3, 2b+3 \tag{7}$$

In other words, we have p=2b+3 and k=4+9/4 (unless it is zero in which the choice of p is invalid).

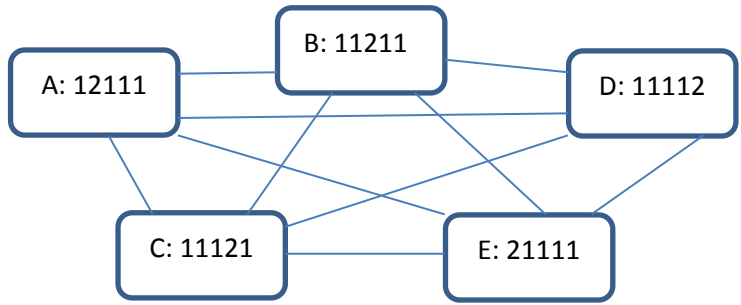

Figure 1. A network using five orthogonal sequences as portals

If we take the sequence to be 1 2 1 1 1, then the 5 shifts of it (12111, 11211, 11121, 11112, 21111) belong to the orthogonal class and each one of these is assigned as the portal key to different users A, B, C, D, and E (Figure 1). The data packet is coded with the key of the intended recipient. Furthermore, for each of these, multiplication with p-1 different constants constitutes further members of the class:

$$\begin{matrix} 1\,2\,1\,1\,1 \\ 2\,4\,2\,2\,2 \\ 3\,2\,3\,3\,3 \\ 4\,1\,4\,4\,4 \\ 5\,3\,5\,5\,5 \\ 6\,5\,6\,6\,6 \end{matrix} \qquad (8)$$

Such multiplication does not change accessibility since the sequence orthogonality class is not affected. The table below gives some example choices that we have for the modulus:

Table 1. Example for equation (8) for M=5

| b | p=2b+3 | k= 4+9/4 mod p |
|---|--------|----------------|
| 2 | 7 | 1 |
| 4 | 11 | 9 |
| 5 | 13 | 3 |
| 7 | 17 | 2 |
| 8 | 19 | 11 |
| 10 | 23 | 12 |
| 13 | 29 | 28 |
| 14 | 31 | 14 |
| 17 | 37 | 34 |

The sequence $a_1, a_2, a_3, ..., a_M$ may be considered a random sequence that is periodic. Owing to (1), it will satisfy the following properties under the condition:

$$\sum_{i=1}^{M} a_i^2 \bmod p = 1 \qquad (9)$$

$$C(k) = \sum_{i=1}^{M} a_i a_{i+k} \bmod p = 0 \text{ for all } k \neq 0 \qquad (10)$$

Property 7 means that the sequence $A(i) = a_1, a_2, a_3, ..., a_M$ may be considered a truly random sequence with autocorrelation function, $C(k)$, that is zero everywhere excepting at k=0.

In other words,

$$\sum A(i)A(i+k) = 0 \text{ for } k \neq 0.$$

**Groups of Orthogonal Sequences**

The sequence A(i) and its M-1 shifts constitute the basic class of orthogonal sequences. It is also clear that w×A is another random sequence in the class for which the autocorrelation function at the origin will be $w^2$ mod p and its value everywhere else will be 0. The peak value of the autocorrelation is obtained from the sequence for which $w^2$ mod p = p-1. Every generator provides us a total of p-1 random sequences.

From this it also follows that

$$\sum [A(i) + A(i+l)]A(i+k) = 0 \text{ for } k \neq 0; k \neq l \qquad (11)$$

This can be used to generate further subclasses. For example, the sequences of (8) together with their cyclic shifts can be put into two orthogonal groups where one group comprises of sequences of rightward shifts through 1 and 2 units and the second group comprises of sequences that have been shifted leftward by 1 and 2 units:

Group I: 12111+11211+11121 = 34443
Group II: 21111+11112= 32223 (12)

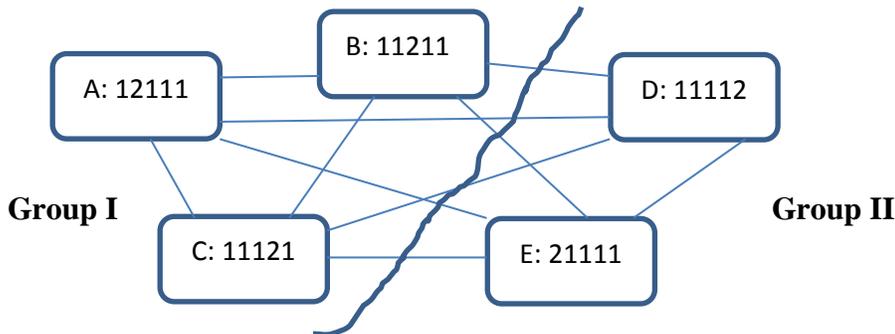

Figure 2. Five users in two subgroups according to equation (12)

Information packets associated with the code 34443 will be accessible to all the members of Class I and inaccessible to members of Class II. The accessibility will be reversed for the access code 32223.

Other equivalent code sequences may be obtained by using different weights for the original sequences. Thus the two groups I and II may also be associated with:

Group I: 12111+3×11211+2×11121 = 60216
Group II: 5×21111+11112= 46660 (12)

It may be checked that the code sequence 60216 is orthogonal to the sequences of D and E and the code sequence 46660 is orthogonal to the sequences of A, B, and C.

*Example 2*. For M=7 and the first row of the circulant matrix as (1,b,1,1,1,1,1), we get the following choices for the modulus and the diagonal term in the product $RR^T=kI$.

Table 2. Choices of circulant matrices for M=7

| b | p= 2b+5 | k=$b^2$+6 mod p |
|---|---------|-----------------|
| 3 | 11 | 4 |
| 4 | 13 | 9 |
| 6 | 17 | 8 |
| 7 | 19 | 17 |
| 9 | 23 | 18 |
| 12 | 29 | 5 |
| 13 | 31 | 20 |
| 18 | 41 | 2 |

The example of Table 2 corresponding to p=41 includes in it a total of 7×40= 280 sequences. The number of sequences equals p×M. Clearly, we can large number of solutions for the modulus. Furthermore, multiplying the matrix with a constant will not change the orthogonality although it will change the modulus. Generalizing we obtain the following result:

**Theorem 1**. Let $a_1, a_2, a_3, ..., a_M$, the first row of the circulant matrix if size M×M be (a, b, a, a, …,a), then

$$RR^T = (M-1)a^2 + b^2 \bmod divisor of\, 2ab + (M-2)a^2 \qquad (13)$$

*Proof.* The proof is a straightforward generalization of construction (7).

*Example 3*. For M=7 and the first row of the circulant matrix as (a,b,a,a,a,a,a), we get the following choices for the modulus and the diagonal term in the product $RR^T=kI$. In the fourth column we always show the largest prime factor that is suitable to be used as modulus. Some cases where the prime modulus is small have not been included in the table.

Table 3. Choices of circulant matrices (a,b,a,a,a,a,a) for M=7

| a | b | 2ab+5$a^2$ | p= suitable prime factor | Diagonal term= 6$a^2$+$b^2$ mod p |
|---|---|-----------|--------------------------|-----------------------------------|
| 2 | 6 | 44 | 11 | 8 |
| 2 | 8 | 52 | 13 | 10 |
| 2 | 12 | 68 | 17 | 15 |
| 2 | 14 | 76 | 19 | 11 |
| 3 | 2 | 57 | 19 | 1 |
| 3 | 4 | 69 | 23 | 1 |
| 3 | 7 | 89 | 89 | 5 |
| 3 | 8 | 93 | 31 | 25 |
| 3 | 11 | 111 | 37 | 27 |
| 4 | 1 | 88 | 11 | 9 |
| 4 | 3 | 104 | 13 | 1 |

Two sequences with the same prime modulus belong to the same as is the case with the sequences in Table 3 with (*a,b*) of (2,14) and (3,2) mod 19, (2,6) and (4,1) mod 11, and (2,8) and (4,3) mod 13. Thus 11×(2,14)=(3,2) mod 19; 2×(2,6)=(4,1) mod 11; and 2×(2,8)=(4,3) mod 13.

*Definition.* Sequences with different generators that are orthogonal for all non-zero shifts will be called product sequences.

**Theorem 2**. Product sequences are transformed from one to another by multiplying by a constant integer.

Now we consider a generating sequence that has three distinct entries.

Let $a_1, a_2, a_3, ..., a_M$, the first row of the circulant matrix if size M×M be (a, b, 1, 1, …,1), then
$$RR^T = a^2 + b^2 + M - 2 \mod divisor(ab + a + b + M - 3, 2a + 2b + M - 4) \quad (11)$$
The product $RR^T$ will now have two distinct non-diagonal terms:

$$a + b + ab + M - 3 \quad \text{and}$$
$$2a + 2b + M - 4 \quad (14)$$

and the diagonal terms will each be $a^2 + b^2 + M - 2$.

For an NHT matrix to exist, we need the following to be true:

$$a + b + ab + M - 3 \equiv 0 \quad 2a + 2b + M - 4 \quad (15)$$

This can be achieved if we pick p = divisor of ($a + b + ab + M - 3, 2a + 2b + M - 4$)

It may be readily seen that the solutions for this case are already contained in (13) if one were to reduce the value of b to its smallest value with respect to the modulus.

This is not the only way to generate random sequences of this type. Other sequences are provided in [17] and [18]. Thus we have for M=7 the set (6, 2, 1, 4, 2, 1, 4) mod 7. Further properties of circulant matrices are provided elsewhere [19].

**Conclusions**
The fact that the NHT-related circulant matrix allows us to generate perfectly random residue sequences opens up the larger question of the property of such random residue sequences. In particular it raises the question whether there is a general algorithm to generate the elements $a_1, a_2, a_3, ..., a_M$. The complexity and scalability of such an algorithm will be of much value in design of certain secure systems.

Orthogonal sequences could have application as keys in environments that are extremely noisy since these strings satisfy certain properties of minimum mutual distance amongst themselves.

**References**

1. X. Wang, Y. Wu and B. Caron, Transmitter identification using embedded pseudo random sequences. IEEE Tran. Broadcasting 50: 244-252 (2004)
2. A. Fuster and L. J. Garcia, An efficient algorithm to generate binary sequences for cryptographic purposes. Theoretical Computer Science 259: 679-688 (2001)
3. S. Kak, The number theoretic Hilbert transform. arXiv:1308.1688
4. S. Kak, The discrete Hilbert transform. Proc. IEEE 58, 585-586 (1970)
5. S.K. Padala and K.M.M. Prabhu, Systolic arrays for the discrete Hilbert transform. Circuits, Devices and Systems, IEE Proceedings 144, 259-264 (1997)
6. F. W. King, Hilbert Transforms. Cambridge University Press (2009)
7. I.G. Roy, On robust estimation of discrete Hilbert transform of noisy data. Geophysics 78, (2013)
8. R. Al-Aifari and A. Katsevich, Spectral analysis of the truncated Hilbert transform with overlap. arXiv:1302.6295
9. I. Noda, Determination of Two-Dimensional Correlation Spectra Using the Hilbert Transform. Applied Spectroscopy, Vol. 54, Issue 7, pp. 994-999 (2000)
10. S. Kak and N.S. Jayant, Speech encryption using waveform scrambling. Bell System Technical Journal 56, 781-808 (1977)
11. N.S. Jayant and S. Kak, Uniform permutation privacy system, US Patent No. 4,100,374, July 11 (1978)
12. S. Kak, Hilbert transformation for discrete data. International Journal of Electronics 34, 177-183 (1973)
13. S. Kak, The discrete finite Hilbert transform. Indian Journal Pure and Applied Mathematics 8, 1385-1390 (1977)
14. S. Kak, Multilayered array computing. Information Sciences 45, 347-365 (1988)
15. S. Kak, A two-layered mesh array for matrix multiplication. Parallel Computing 6, 383-385 (1988)
16. R. Kandregula, The basic discrete Hilbert transform with an Information Hiding Application. 2009. arXiv:0907.4176
17. V.K. Kotagiri, The number theoretic Hilbert transform. http://www.cs.okstate.edu/~subhashk/nht2kotagiri.pdf
18. V.K. Kotagiri, New results on the number theoretic Hilbert transform. arXiv:1310.6924
19. S. Kak, Notes on NHT-circulant matrices. http://www.cs.okstate.edu/~subhashk/nhtcirculant.pdf